\documentclass[10pt,twocolumn]{article}

\usepackage[T1]{fontenc}
\usepackage[utf8]{inputenc}
\usepackage{mathptmx}                   % Classic Times Roman (IEEE Style)
\usepackage[scaled=0.92]{helvet}        % Clean Helvetica for sans
\usepackage{courier}                    % Fixed-width Courier
\usepackage{microtype}                  % Optimal kerning and character protrusion
\usepackage{amsmath,amssymb}
\usepackage{booktabs}
\usepackage{array}
\usepackage{tabularx}
\usepackage{multirow}
\usepackage{makecell}
\usepackage{graphicx}
\usepackage[table]{xcolor}
\usepackage{tikz}
\usetikzlibrary{trees,positioning,arrows.meta,calc}
\usepackage{listings}
\usepackage{enumitem}
\usepackage{float}
\usepackage{caption}
\usepackage{geometry}
\usepackage{fancyhdr}
\usepackage{titlesec}
\usepackage{dblfloatfix}                 % Proper bottom/top placement of double-column floats
\usepackage{environ}
\usepackage{cuted}
\usepackage{accsupp}                   % Anti-bot PDF text layer obfuscation
\usepackage{balance}                   % Balanced two-column bibliography on final page

\setlist{topsep=2.5pt plus 1.5pt minus 1.0pt, itemsep=1.5pt plus 1.2pt minus 0.8pt, partopsep=0pt, parsep=0pt, leftmargin=1.2em}
\AtBeginDocument{%
  \setlength{\abovedisplayskip}{3.5pt plus 2.5pt minus 1.5pt}%
  \setlength{\belowdisplayskip}{3.5pt plus 2.5pt minus 1.5pt}%
  \setlength{\abovedisplayshortskip}{1.5pt plus 1.5pt minus 1.0pt}%
  \setlength{\belowdisplayshortskip}{2.5pt plus 1.5pt minus 1.0pt}%
}

\definecolor{paperdark}{HTML}{111827}
\definecolor{papergray}{HTML}{4B5563}
\definecolor{paperlight}{HTML}{F9FAFB}
\definecolor{codeborder}{HTML}{E5E7EB}
\definecolor{codegreen}{HTML}{15803D}
\definecolor{codegray}{HTML}{6B7280}
\definecolor{codered}{HTML}{B91C1C}

\colorlet{paperblue}{paperdark}
\colorlet{lqblue}{paperdark}
\colorlet{lqdark}{paperdark}
\colorlet{lqgray}{papergray}
\definecolor{lqlight}{HTML}{F3F4F6}

\usepackage{xurl}
\usepackage[hidelinks,colorlinks=true,
            linkcolor=paperdark,
            urlcolor=paperdark,
            citecolor=paperdark]{hyperref}

\titleformat{\section}
  {\normalsize\scshape\bfseries\color{paperdark}\centering}{\Roman{section}.}{0.5em}{}
\titleformat{\subsection}
  {\small\bfseries\color{paperdark}}{\Alph{subsection}.}{0.5em}{}
\titleformat{\subsubsection}
  {\small\itshape\color{papergray}}{\arabic{subsubsection})}{0.5em}{}

\titlespacing*{\section}{0pt}{8pt plus 4pt minus 2pt}{3pt plus 2pt minus 1pt}
\titlespacing*{\subsection}{0pt}{6pt plus 3pt minus 2pt}{2.5pt plus 1.5pt minus 1pt}
\titlespacing*{\subsubsection}{0pt}{4pt plus 2pt minus 1pt}{2pt plus 1pt minus 0.5pt}

\renewcommand{\headrulewidth}{0.3pt}
\renewcommand{\headrule}{\hbox to\headwidth{\color{codeborder}\leaders\hrule height \headrulewidth\hfill}}

\renewenvironment{abstract}
  {\small\noindent\textbf{\textit{Abstract}}---\ignorespaces}
  {\par\vspace{6pt}}

\newcommand{\bitfield}[1]{\texttt{#1}}
\newcommand{\msgtype}[1]{\textsc{#1}}
\newcommand{\callsign}[1]{\texttt{#1}}
\newcommand{\LQ}{LQ}

\newcommand{\szero}{\leavevmode\hbox{\tikz[baseline=-0.55ex]{%
  \node[inner sep=0pt,outer sep=0pt] (n) {\texttt{0}};%
  \draw[line width=0.38pt] ([shift={(-0.11em,-0.19em)}]n.center) -- ([shift={(0.11em,0.19em)}]n.center);%
}}}

\DeclareRobustCommand{\lightning}{\leavevmode\hbox{\tikz[baseline=-0.35ex,scale=0.75,line join=round,line cap=round]{%
  \fill (0.06em,0.52em) -- (-0.16em,0.06em) -- (0.02em,0.06em) -- (-0.08em,-0.50em) -- (0.18em,-0.02em) -- (0.00em,-0.02em) -- (0.14em,0.52em) -- cycle;%
}}}
\DeclareUnicodeCharacter{2301}{\ensuremath{\lightning}}
\DeclareUnicodeCharacter{26A1}{\ensuremath{\lightning}}
\DeclareUnicodeCharacter{21AF}{\ensuremath{\lightning}}
\DeclareUnicodeCharacter{2404}{\texttt{[EOM]}}

\begin{document}

\begin{strip}
% ---- IEEE Title Block ------------------------------------------------------
\begin{center}
  \vspace*{-12pt}
  {\LARGE\bfseries\color{paperdark}
    The LQ Digital Mode Family:\\[3pt]
    Protocol Architecture and Reference Specification\\[3pt]
    for Weak-Signal Communications\par}
  \vspace{10pt}
  {\large Luis Quesada (HB9IPH) --- Independent Researcher\par}
  \vspace{3pt}
  {\small\texttt{\BeginAccSupp{ActualText={l-u-i-s [at] l-q-8 [dot] o-r-g}}luis@lq8.org\EndAccSupp{}}\par}
  \vspace{8pt}
\end{center}

\begingroup
  \let\origclearpage\clearpage
  \renewcommand{\clearpage}{\par\vspace{4pt}}

\begin{abstract}
\noindent
This paper introduces \LQ8, an open-source weak-signal digital mode for amateur radio that completes structured two-way contacts in 4 transmissions (60\,s), achieving a $1.50\times$ speedup over canonical 6-transmission exchanges (90\,s) and $1.25\times$ over 5-transmission \texttt{RR73} exchanges (75\,s) while preserving complete bidirectional exchange of Maidenhead grid locators, signal reports, and mutual acknowledgments.
By replacing fixed-width message-type headers with variable-length prefix codes matched to physical-layer bit budgets, \LQ8 packs the answering station's grid locator and measured signal report into a single 77-bit reply.
In multi-station pileup operations, \LQ8 confirms up to two answering stations simultaneously within a single-carrier 50.0\,Hz transmission with a 0.0\,dB RF power-splitting penalty, achieving a peak rate of 240\,QSOs/h in continuous pileup ($4\times$ over single-carrier FT8; 480\,QSOs/h dual-carrier) while retaining $-21.0$\,dB SNR sensitivity.
The protocol family also defines three extended profiles (\LQ16, \LQ4, and \LQ2) tailored for extreme sensitivity, VHF/UHF contesting, and fast-burst channels.
\end{abstract}

\endgroup
\vspace{-6pt}
\end{strip}

\section{Introduction}
\label{sec:intro}

The FT8 digital mode~\cite{ft8}, designed by Steve Franke (K9AN) and Joe Taylor (K1JT), has transformed amateur radio since its
introduction in 2017 in the WSJT-X software suite~\cite{wsjtx}, enabling reliable contacts at signal-to-noise ratios
as low as $-21.0$\,dB in a 2500\,Hz bandwidth.  Its 77-bit message payload,
combined with a low-density parity-check (LDPC) forward error correction
code~\cite{gallager} and 8-tone Gaussian frequency-shift keying (8-GFSK), provides a
robust physical layer that is now well-proven across millions of contacts
worldwide.

In conventional weak-signal protocols (e.g., FT8), a structured two-way contact canonically requires six sequential transmissions across 15-second time slots (90 seconds total), or five transmissions (75 seconds) when using shortened \texttt{RR73} confirmations (\msgtype{CQ} $\to$ \msgtype{CALL} $\to$ \msgtype{REPORT} $\to$ \msgtype{RREPORT} $\to$ \msgtype{RR73}). Under a 77-bit payload budget, encoding two 28-bit standard callsigns, a 15-bit Maidenhead grid locator, and a 5-bit signal report requires 76 information bits ($28+28+15+5=76$). Uniform fixed-width type headers (typically 3 or 4 bits) exceed the remaining 1-bit margin, preventing the simultaneous transmission of locator and report in a single transmission.

\textbf{LQ8} resolves this constraint by employing \textbf{variable-length prefix codes}~\cite{huffman} tailored to physical-layer bit budgets. By allocating a 1-bit prefix to the dominant response message format, \LQ8 preserves 76 bits for payload data, packing both the answering station's grid locator and measured signal report into a single frame. This completes confirmed contacts in 4 transmissions (60 seconds) with full mutual verification, while retaining the proven continuous-phase 8-GFSK and LDPC$(174,91)$ physical layer ($-21.0$\,dB SNR sensitivity). Furthermore, for high-rate DXpedition Fox and Hound (F/H) pileups, \LQ8 introduces the \msgtype{MULTI-REPORT+73} protocol to confirm two answering Hound stations simultaneously within a single-carrier transmission without RF power-splitting penalties. Cross-isolated Costas synchronization arrays ($C_7 = [2, 5, 6, 1, 3, 0, 4]$, exhibiting $\le 1$ coincidence hit) isolate traffic from legacy decoders. Physical layer, CRC-14, and LDPC parameters are presented in Section~\ref{sec:phy}.

Building upon \LQ8, the protocol family defines three complementary profiles (\LQ16, \LQ4, and \LQ2) tailored for ultra-weak fading channels, rapid contesting, and fast burst links.

\subsection{Architectural Philosophy \& Design Principles}
\label{sec:advantages}

The \LQ{} family is engineered around four core design tenets:
\begin{itemize}[leftmargin=1.5em, itemsep=1.5pt plus 1.0pt minus 0.5pt, topsep=2.0pt plus 1.0pt minus 0.5pt]
  \item \textbf{Payload-Matched Prefix Allocation:} Replacing uniform headers with variable-length prefix codes frees capacity for simultaneous locator and report transmission, shortening 1-on-1 contacts to 4 slots (60\,s) and reducing vulnerability to ionospheric fading (QSB) and transient interference.
  \item \textbf{Coexistence and Spectrum Preservation:} All modes maintain continuous-phase GFSK and LDPC$(174,91)$ channel coding while employing low-cross-coincidence Costas sync arrays to prevent cross-protocol false decodes.
  \item \textbf{Robust Special-Case Support:} Non-standard callsigns (up to 13 characters), dedicated 1-bit portable suffixes (\texttt{0}=\emph{none}, \texttt{1}=\texttt{/P}), 24/20-bit collision-resistant callsign hashes, and a 104-character free-text Varicode provide seamless support for contesting, portable activations (POTA/SOTA), and DXpeditions.
  \item \textbf{Clear-Text Station Identification:} Transmitting stations convey their callsigns in clear text across all 1-on-1 exchanges (including non-standard caller replies via 48-bit Base-38 encoding), reserving compact hashes strictly for target addressing and multi-station pileup frames.
\end{itemize}

\subsection{Implementations \& Reference Software}
\label{sec:implementations}

The complete suite is implemented in the open-source C++ reference library (\texttt{lq\_lib})~\cite{lqlib} (MIT License), featuring an automated verification harness and golden vector test suites across all four modulation profiles. Web demonstrations and interactive protocol resources are published on the official project portal~\cite{lqportal}. Native portable operation is provided by \textbf{qFT8}~\cite{qft8}, deploying \LQ8 alongside live testbeds for \LQ16, \LQ4, and \LQ2.

\pagebreak

\section{QSO Protocol}
\label{sec:qso}

The \LQ{} protocol establishes deterministic state transitions, transmission slot timing, and multi-station arbitration to achieve confirmed bilateral contacts across low-SNR channels.

\subsection{Standard Four-Message Exchange}
\label{sec:qso:standard}

The \LQ{} QSO protocol is built upon two core principles: bidirectional exchange of Maidenhead grid locators and signal reports (measured SNR in dB), coupled with mutual confirmation (Table~\ref{tab:ft8qso}). Following weak-signal convention, \LQ{} establishes an operational termination criterion: each station must hear an explicit confirmation after delivering its own signal report before considering the contact complete. Station~A considers the contact complete upon receiving Station~B's concluding \msgtype{73} in T4, while Station~B considers it complete upon decoding Station~A's report in T3 and transmitting its final \msgtype{73}.

\begin{table}[htbp]
\centering
\caption{QSO sequence comparison between conventional protocols (FT8) and the \LQ8 4-slot protocol.}
\label{tab:ft8qso}
\resizebox{\columnwidth}{!}{%
\begin{tabular}{@{}cl|cl@{}}
\toprule
\textbf{Seq} & \textbf{Conventional Protocols (FT8)} & \textbf{Seq} & \textbf{\LQ8 Protocol} \\
\midrule
1 & \texttt{CQ YO1YO JN47}                     & 1   & \texttt{CQ YO1YO JN47} \\
2 & \texttt{YO1YO TU2TU KL22}                  & 2   & \texttt{YO1YO TU2TU KL22 -03} \\
3 & \texttt{TU2TU YO1YO +05}                   & 3   & \texttt{TU2TU YO1YO R+05} \\
4 & \texttt{YO1YO TU2TU R-03}                  & 4   & \texttt{YO1YO TU2TU 73} \emph{[Confirmed]} \\
5 & \texttt{TU2TU YO1YO RR73} (or \texttt{RRR})& --- & --- \\
6 & \texttt{YO1YO TU2TU 73} (optional if \texttt{RR73}) & --- & --- \\
\bottomrule
\end{tabular}%
}
\end{table}

As summarized in Table~\ref{tab:ft8qso}, \LQ8 completes a contact in 4 time slots (60\,s vs.\ 5--6 messages / 75--90\,s in conventional exchanges ending in \texttt{RR73} or \texttt{RRR}/73), delivering full bidirectional callsigns, Maidenhead locators, and signal reports with mutual verification:
\begin{enumerate}[leftmargin=1.4em, itemsep=0.8pt plus 1.0pt minus 0.4pt, topsep=1.5pt plus 1.0pt minus 0.5pt, parsep=0pt]
  \item \textbf{T1 (Slot 1, Even):} Station~A transmits a \msgtype{CQ} broadcast (catalogued in Section~\ref{sec:msg_arch}, Types~1--4) containing its callsign, optional portable suffix, grid locator, or CQ modifier.
  \item \textbf{T2 (Slot 2, Odd):} Station~B transmits a directed \msgtype{CALL} reply (Types~5--7) packing Station~A's target address, caller callsign, measured SNR report, and grid locator into a single frame.
  \item \textbf{T3 (Slot 3, Even):} Station~A acknowledges with \msgtype{REPORT+73} (Type~8 for standard callsigns, or multi-station Type~11), delivering its measured SNR report for Station~B.
  \item \textbf{T4 (Slot 4, Odd):} Station~B transmits a final \msgtype{73} acknowledgment (Types~9 or 10; Type~12 for 10--13-character compound callsigns), completing the confirmed bidirectional contact.
\end{enumerate}

\subsection{Retransmission, Error Recovery and Logging Semantics}
\label{sec:qso:retx}

To ensure session progress over lossy channels, \LQ{} formalizes timeout recovery and logging rules.

\paragraph{Retransmission and Repeat Calls.}
If Station~B misses \msgtype{REPORT+73} in T3 (due to QSB or interference), it repeats \msgtype{CALL} in the next odd slot. Receiving this repeat informs Station~A that its report was lost, prompting Station~A to resend \msgtype{REPORT+73}. Similarly, if Station~A misses the final \msgtype{73} in T4, it retransmits \msgtype{REPORT+73}; Station~B re-acknowledges with \msgtype{73} without duplicate logging.

\paragraph{QSO Logging Semantics.}
Contacts are logged once callsigns, locators, reports, and acknowledgments are exchanged: Station~A logs upon decoding the final \msgtype{73} in T4 (or transmitting \msgtype{REPORT+73} in Fox pileup mode); Station~B logs upon transmitting \msgtype{73} in T4 (having decoded Station~A's report in T3). In ADIF logging (LoTW, QRZ.com, Club Log), contacts map to mode \texttt{DATA} with submode \texttt{LQ8} (\texttt{LQ4}, \texttt{LQ2}, \texttt{LQ16}). A watchdog aborts after three unacknowledged attempts (45\,s in \LQ8, 22.5\,s in \LQ4), returning to idle monitoring.

\subsection{DXpedition Fox and Hound Multi-Station Pileup Operation}
\label{sec:qso:pileup}

During DXpedition pileups or contest operations in Fox and Hound (F/H) mode, Station~A (Fox) activates the \msgtype{MULTI-REPORT+73} protocol (Type~11) to confirm up to two answering Hound stations simultaneously on a single carrier:
\begin{enumerate}[leftmargin=1.4em, itemsep=0.8pt plus 1.0pt minus 0.4pt, topsep=1.5pt plus 1.0pt minus 0.5pt, parsep=0pt]
  \item \textbf{Slot 1 (Fox):} Fox broadcasts directed \msgtype{CQ} on its primary run frequency (e.g., \texttt{CQ DX FOX}).
  \item \textbf{Slot 2 (Hounds):} Remote Hounds call simultaneously on split audio frequencies via directed \msgtype{CALL} (e.g., \texttt{<FOX> <HOUND1> FN42 -08}).
  \item \textbf{Slot 3 (Fox):} Fox confirms both Hounds simultaneously on the run frequency via \msgtype{MULTI-REPORT+73} (\texttt{<HOUND1> R+05 <HOUND2> R-02 <FOX>}).
  \item \textbf{Slot 4 (Hounds):} Both Hounds respond with concluding \msgtype{73} acknowledgments on their split frequencies (\texttt{<FOX> <HOUND1> 73}).
\end{enumerate}
When confirming a single Hound in Fox mode, standard callers are confirmed using Type~8 (\msgtype{REPORT+73 std+suf}). Single-target Type~11 (or Type~12 for final acknowledgments) is used when confirming non-standard callers or when operating under strict split-frequency F/H scheduling; in this case, Target~2 duplicates Target~1 (identical hash, and identical SNR in Type~11), which decoders process as a single-station transmission.

\paragraph{Throughput Multiplier Analysis.}
In continuous pileups where Step~3 overlaps with incoming calls, completing 2 QSOs every 2 slots (30\,s) delivers a peak rate of 240\,QSOs/h ($4\times$ over single-carrier FT8; 480\,QSOs/h dual-carrier; 160\,QSOs/h in isolated 3-slot exchanges) with \textbf{0.0\,dB power penalty}, avoiding the $-3.0$\,dB RF loss of multi-carrier FT8 while halving fading exposure.

\subsection{Multi-Caller Contention and Pileup Queuing}
\label{sec:qso:contention}

When multiple stations answer the same CQ simultaneously in 1-on-1 mode, Station~A selects one caller and transmits \msgtype{REPORT+73} addressed to that station; other callers retry in the next alternating time slot. Frames are disambiguated via uniform 24-bit callsign hashes ($2^{24} \approx 16.78\times 10^6$ bins, Section~\ref{sec:hash}) and canonical precedence (\mbox{Section~\ref{sec:msg:limitations}}).

To optimize pileup resolution, transceivers implement opportunistic candidate queuing (up to 16 stations). Priority aging balances throughput with fairness, preventing high-SNR arrivals from starving marginal callers. Upon decoding \msgtype{73} in T4, Station~A immediately directs \msgtype{REPORT+73} to the highest-priority queued station without an intermediate \msgtype{CQ}, accelerating turnover. If the queue is full, lowest-SNR entries below decode threshold are evicted first; entries expire after four unacknowledged cycles (60\,s in \LQ8, 30\,s in \LQ4).

\pagebreak

\section{Message Architecture}
\label{sec:msg_arch}
\label{sec:typedetails}
\label{sec:huffman}

To maximize channel efficiency within the 77-bit physical frame budget, \LQ{} replaces fixed-width bit fields with an entropy-matched variable-length message architecture.

\subsection{Design Motivation \& Prefix-Free Framing}
\label{sec:huffman:motivation}

Under a strict 77-bit physical payload budget, conventional fixed-width message headers impose an unnecessary penalty on high-entropy transmissions. In a standard \msgtype{CALL} reply, encoding two 28-bit standard callsigns, a 15-bit Maidenhead grid locator, and a 5-bit measured SNR report requires 76 information bits:
\begin{equation}
  \underbrace{1\text{b}}_{\text{Prefix}} + \underbrace{28\text{b}}_{\text{Target}} + \underbrace{28\text{b}}_{\text{Caller}} + \underbrace{15\text{b}}_{\text{Locator}} + \underbrace{5\text{b}}_{\text{SNR}} = \mathbf{77\text{ bits}}.
\label{eq:call_budget}
\end{equation}
Uniform 3-bit or 4-bit type headers exceed the remaining 1-bit margin, preventing simultaneous transmission of locator and report in conventional protocols.

\LQ{} resolves this constraint by synthesizing variable-length prefix coding~\cite{huffman,shannon} with physical-layer bit budgets. Assigning a \textbf{1-bit} codeword (\texttt{1}) to the dominant \msgtype{CALL std (nosuf)} format preserves 76 bits for payload data, combining station identification, locator delivery, and signal report measurement into a single frame. Simpler messages (such as standard CQ, requiring 43 payload bits) readily accommodate longer prefix codes.

Formally, the prefix code $\mathcal{C} = \{c_1, c_2, \ldots, c_k\}$ is \emph{prefix-free}:
\begin{equation}
  \forall\, i \neq j: \quad c_i \text{ is not a prefix of } c_j.
\end{equation}
The receiver traverses a binary prefix tree bit-by-bit from the most-significant bit until reaching a terminal leaf node, identifying the message type without length fields or delimiters. Although the mathematical binary prefix tree across all 16 slots satisfies $\sum_{i=1}^{16} 2^{-\text{len}_i} = 1.0$, the active operational codebook is intentionally prefix-free without being Kraft-complete because Types~14--16 are explicitly reserved for future backward-compatible protocol expansions rather than active operational message formats.

\subsection{Operational Categories \& Message Catalogue}
\label{sec:msg:catalogue}

Stations select transmission formats according to the operational categories defined in Table~\ref{tab:categories_of_operation}. Table~\ref{tab:huffman} catalogues all 16 message slots with their prefix codewords, payload layouts, and bit budgets. Frame payloads $\le 77$ bits are zero-padded to the 77-bit link-layer boundary. Payloads are serialized in big-endian bit order (MSB first) into a 10-byte buffer: bit 0 corresponds to \texttt{0x80} of Byte 0, bit 7 to \texttt{0x01} of Byte 0, and bits 77--79 (the three least significant bits of Byte 9) are set to zero.

\textbf{Field Designations in Table~\ref{tab:huffman}:}
\begin{itemize}[leftmargin=1.4em, itemsep=1.5pt plus 1.0pt minus 0.5pt, topsep=2pt plus 1.0pt minus 0.5pt, parsep=0pt]
  \item \textbf{Canonical Short Identifiers:} Standardized mnemonics: \textbf{CQ}, \textbf{CALL}, \textbf{RPT73} (for \msgtype{REPORT+73}), \textbf{73}, \textbf{RPT73M} (for \msgtype{MULTI-REPORT+73}), and \textbf{73M} (for \msgtype{MULTI-73}).
  \item \textbf{Prefix \& Standard Callsign:} ``Code'' denotes prefix length; ``call:std'' is a 28-bit standard callsign (Section~\ref{sec:std_call}).
  \item \textbf{Non-Standard Callsigns \& Suffixes:} ``call:$n$'' is an $n$-character non-standard callsign packed via Base-38 (Section~\ref{sec:nonstd_call}); ``Suf'' encodes 1-bit portable modifiers (\texttt{0}=\emph{none}, \texttt{1}=\texttt{/P}). Type~5 omits suffixes to pack two standard callsigns, locator, and SNR into 77 bits ($1+28+28+15+5=77\text{b}$).
  \item \textbf{Callsign Hashes:} ``Hash'' is a 24-bit CRC-24/Q hash ($2^{24} \approx 16.78\times 10^6$ bins); ``hash$^\dagger$'' is the 16-bit DX hash (Section~\ref{sec:hash}).
  \item \textbf{Multi-Station Fields:} Type~11 Fox pileup sequences $[\text{Prefix}_3 \mid \text{DX}_{16} \mid \text{Tgt1}_{24} \mid \text{SNR1}_5 \mid \text{Tgt2}_{24} \mid \text{SNR2}_5]$ (77 bits); Type~12 formats $[\text{Prefix}_8 \mid \text{DX}_{16} \mid \text{Tgt1}_{24} \mid \text{Tgt2}_{24} \mid 00000_2]$ (77 bits, 5 zero-padded bits).
  \item \textbf{Modifiers, Locators \& SNR:} ``Mod'' is a 20-bit modifier (Section~\ref{sec:modifier}); ``loc'' is a 15-bit locator (Section~\ref{sec:locator}); ``SNR'' is a 5-bit report (Section~\ref{sec:rst}).
\end{itemize}

\begin{table*}[p]
\centering

% --- Table 2: Operational Categories ---
\small
\setlength{\tabcolsep}{4pt}
\renewcommand{\arraystretch}{0.95}
\caption{Operational categories and corresponding 4-step message sequences.}
\label{tab:categories_of_operation}
\begin{tabular*}{\textwidth}{@{\extracolsep{\fill}}clcccc@{}}
\toprule
\textbf{Cat.} & \textbf{Callsign / Mode Configuration} & \textbf{Step 1: CQ} & \textbf{Step 2: CALL} & \textbf{Step 3: REPORT+73} & \textbf{Step 4: 73} \\
\midrule
\textbf{A} & Standard (no suffixes) & Type 1 & Type 5 & Type 8 & Type 9 \\
\textbf{B} & Standard + std suffix (\texttt{/P}) & Type 1 & Type 6 & Type 8 & Type 9 \\
\textbf{C} & Non-standard ($\le 9$ chars) + optional suffix & Type 2 / 3 & Type 7$^*$ / Type 6 & Type 11 & Type 10 \\
\textbf{D} & Long compound ($10$--$13$ chars) + optional suffix & Type 4 & ---$^\ddagger$ & Type 11 & Type 12 \\
\textbf{E} & Multi-station reply & --- & --- & Type 11 (Fox) & Type 9 / 10 (Hounds)$^\S$ \\
\bottomrule
\multicolumn{6}{@{}p{\textwidth}@{}}{\vspace{2pt}\footnotesize $^*$In Step~2, Category~C callers transmit Type~7 (clear-text callsign, suffix, SNR to target 20-bit hash); standard callers answering a Category~C CQ transmit Type~6 (locator, SNR to target 24-bit hash). \quad $^\ddagger$Category~D compound stations ($10$--$13$ chars) initiate contact via CQ (Type~4) or free text (Type~13). \quad $^\S$A Hound with a 10--13-character callsign sends Type~12; the Fox may send an optional \msgtype{MULTI-73} (Type~12) after the exchange.}
\end{tabular*}

\vspace{6pt plus 1fill}

% --- Table 3: Complete Message Catalogue ---
\caption{Complete \LQ{} message-type catalogue.}
\label{tab:huffman}
\small
\setlength{\tabcolsep}{2.5pt}
\renewcommand{\arraystretch}{0.94}
\begin{tabular*}{\textwidth}{@{\extracolsep{\fill}}lclccccccccr@{}}
\toprule
\textbf{Huffman Code} & \textbf{ID} & \textbf{Type} & \multicolumn{8}{c}{\textbf{Bits (Payload)}} & \textbf{Total} \\
\cmidrule(lr){4-11}
& & & \textbf{Code} & \textbf{Target} & \textbf{Suf} & \textbf{Caller} & \textbf{Suf} & \textbf{Modifier} & \textbf{Locator} & \textbf{SNR} & \\
\midrule
\bitfield{\szero\szero\szero\szero\szero\szero\szero\szero\szero\szero}         & 1  & CQ std+suf                      & 10 & ---           & ---   & 28 call:std   & 1 suf & 20 mod & 15 loc & ---   & 74 \\
\bitfield{\szero\szero\szero\szero\szero\szero\szero\szero\szero1}              & 2  & CQ non-std~1                    & 10 & ---           & ---   & 48 call:9     & 1 suf & ---    & 15 loc & ---   & 74 \\
\bitfield{\szero\szero\szero\szero\szero\szero\szero1}                          & 3  & CQ non-std~2                    & 8  & ---           & ---   & 48 call:9     & 1 suf & 20 mod & ---    & ---   & 77 \\
\bitfield{\szero\szero\szero\szero\szero\szero1}                                  & 4  & CQ non-std~3                    & 7  & ---           & ---   & 69 call:13    & 1 suf & ---    & ---    & ---   & 77 \\
\bitfield{1}                                                                  & 5  & CALL std (nosuf)                & 1  & 28 call:std   & ---$^*$ & 28 call:std & ---$^*$ & ---  & 15 loc & 5 SNR & 77 \\
\bitfield{\szero1\szero\szero}                                                & 6  & CALL std+suf                    & 4  & 24 hash       & ---   & 28 call:std   & 1 suf & ---    & 15 loc & 5 SNR & 77 \\
\bitfield{\szero\szero1}                                                       & 7  & CALL non-std                    & 3  & 20 hash       & ---   & 48 call:9     & 1 suf & ---    & ---    & 5 SNR & 77 \\
\bitfield{\szero\szero\szero\szero\szero\szero\szero\szero1\szero}              & 8  & REPORT+73 std+suf (RPT73)       & 10 & 28 call:std   & 1 suf & 28 call:std   & 1 suf & ---    & ---    & 5 SNR & 73 \\
\bitfield{\szero\szero\szero\szero\szero\szero\szero\szero11}                  & 9  & 73 std+suf                      & 10 & 28 call:std   & 1 suf & 28 call:std   & 1 suf & ---    & ---    & ---   & 68 \\
\bitfield{\szero\szero\szero1}                                                & 10 & 73 non-std                      & 4  & 24 hash       & ---   & 48 call:9     & 1 suf & ---    & ---    & ---   & 77 \\
\bitfield{\szero11}                                                           & 11 & MULTI-REPORT+73 (RPT73M)        & 3  & \multicolumn{7}{c}{16 hash$^\dagger$ \quad+\quad 2 $\times$ (24 hash + 5 SNR)} & 77 \\
\bitfield{\szero\szero\szero\szero\szero111}                                  & 12 & MULTI-73 (73M)                  & 8  & \multicolumn{7}{c}{16 hash$^\dagger$ \quad+\quad 2 $\times$ 24 hash}            & 72 \\
\bitfield{\szero1\szero1}                                                     & 13 & FREE TEXT                       & 4  & \multicolumn{7}{c}{73 text (Varicode)} & 77 \\
\bitfield{\szero\szero\szero\szero1}                                           & 14 & \emph{(reserved A)}             & 5  & \multicolumn{7}{c}{72 payload} & 77 \\
\bitfield{\szero\szero\szero\szero\szero1\szero}                              & 15 & \emph{(reserved B)}             & 7  & \multicolumn{7}{c}{70 payload} & 77 \\
\bitfield{\szero\szero\szero\szero\szero11\szero}                             & 16 & \emph{(reserved C)}             & 8  & \multicolumn{7}{c}{69 payload} & 77 \\
\bottomrule
\multicolumn{12}{@{}p{\textwidth}@{}}{\vspace{2pt}\footnotesize $^*$Type~5 omits suffix fields to pack two standard callsigns, locator, and SNR into 77 bits ($1+28+28+15+5=77\text{b}$). \quad $^\dagger$16-bit DX hash (Section~\ref{sec:hash}). \quad The ``Total'' column indicates active information bits; formats with Total $< 77$ bits (Types 1, 2, 8, 9, 12) are right-padded with trailing zeros ($77 - \text{Total}$ bits) to complete the 77-bit physical frame boundary prior to CRC-14 computation.}
\end{tabular*}

\vspace{6pt plus 1fill}

\caption{LQ Varicode assignments for the 104-character free-text alphabet.}
\label{tab:varicode}
\resizebox{\textwidth}{!}{%
\scriptsize
\setlength{\tabcolsep}{1.2pt}
\begin{tabular}{@{}cl c r | cl c r | cl c r | cl c r@{}}
\toprule
\textbf{Char} & \textbf{Varicode} & \textbf{Len} & \textbf{Freq} & \textbf{Char} & \textbf{Varicode} & \textbf{Len} & \textbf{Freq} & \textbf{Char} & \textbf{Varicode} & \textbf{Len} & \textbf{Freq} & 
\textbf{Char} & \textbf{Varicode} & \textbf{Len} & \textbf{Freq} \\
\midrule
\texttt{{\textvisiblespace}} & \texttt{\szero{}\szero{}1} & 3 & 16.73\% & \texttt{1} & \texttt{\szero{}\szero{}\szero{}\szero{}\szero{}\szero{}1\szero{}11} & 10 & 0.09\% & \texttt{*} & \texttt{\szero{}\szero{}\szero{}\szero{}\szero{}\szero{}\szero{}\szero{}\szero{}\szero{}11\szero{}} & 13 & 0.01\% & \texttt{Å} & \texttt{\szero{}\szero{}\szero{}\szero{}\szero{}\szero{}\szero{}\szero{}\szero{}\szero{}\szero{}\szero{}1\szero{}111} & 17 & $<$0.01\% \\
\texttt{E} & \texttt{\szero{}1\szero{}} & 3 & 10.23\% & \texttt{Q} & \texttt{\szero{}\szero{}\szero{}\szero{}\szero{}\szero{}11\szero{}\szero{}} & 10 & 0.07\% & \texttt{[} & \texttt{\szero{}\szero{}\szero{}\szero{}\szero{}\szero{}\szero{}\szero{}\szero{}\szero{}111\szero{}} & 14 & 0.01\% & \texttt{Æ} & \texttt{\szero{}\szero{}\szero{}\szero{}\szero{}\szero{}\szero{}\szero{}\szero{}\szero{}\szero{}\szero{}11\szero{}\szero{}\szero{}} & 17 & $<$0.01\% \\
\texttt{T} & \texttt{\szero{}11\szero{}} & 4 & 7.44\% & \texttt{!} & \texttt{\szero{}\szero{}\szero{}\szero{}\szero{}\szero{}11\szero{}1} & 10 & 0.07\% & \texttt{]} & \texttt{\szero{}\szero{}\szero{}\szero{}\szero{}\szero{}\szero{}\szero{}\szero{}\szero{}1111} & 14 & 0.01\% & \texttt{Ç} & \texttt{\szero{}\szero{}\szero{}\szero{}\szero{}\szero{}\szero{}\szero{}\szero{}\szero{}\szero{}\szero{}11\szero{}\szero{}1} & 17 & $<$0.01\% \\
\texttt{A} & \texttt{\szero{}111} & 4 & 6.59\% & \texttt{?} & \texttt{\szero{}\szero{}\szero{}\szero{}\szero{}\szero{}111\szero{}} & 10 & 0.07\% & \texttt{$|$} & \texttt{\szero{}\szero{}\szero{}\szero{}\szero{}\szero{}\szero{}\szero{}\szero{}\szero{}\szero{}1\szero{}\szero{}} & 14 & 0.01\% & \texttt{È} & \texttt{\szero{}\szero{}\szero{}\szero{}\szero{}\szero{}\szero{}\szero{}\szero{}\szero{}\szero{}\szero{}11\szero{}1\szero{}} & 17 & $<$0.01\% \\
\texttt{O} & \texttt{1\szero{}\szero{}\szero{}} & 4 & 6.24\% & \texttt{Z} & \texttt{\szero{}\szero{}\szero{}\szero{}\szero{}\szero{}1111} & 10 & 0.06\% & \texttt{/} & \texttt{\szero{}\szero{}\szero{}\szero{}\szero{}\szero{}\szero{}\szero{}\szero{}\szero{}\szero{}1\szero{}1\szero{}} & 15 & $<$0.01\% & \texttt{É} & \texttt{\szero{}\szero{}\szero{}\szero{}\szero{}\szero{}\szero{}\szero{}\szero{}\szero{}\szero{}\szero{}11\szero{}11} & 17 & $<$0.01\% \\
\texttt{N} & \texttt{1\szero{}\szero{}1} & 4 & 5.96\% & \texttt{,} & \texttt{\szero{}\szero{}\szero{}\szero{}\szero{}\szero{}\szero{}1\szero{}\szero{}\szero{}} & 11 & 0.06\% & \texttt{\$} & \texttt{\szero{}\szero{}\szero{}\szero{}\szero{}\szero{}\szero{}\szero{}\szero{}\szero{}\szero{}1\szero{}11\szero{}} & 16 & $<$0.01\% & \texttt{Ê} & \texttt{\szero{}\szero{}\szero{}\szero{}\szero{}\szero{}\szero{}\szero{}\szero{}\szero{}\szero{}\szero{}111\szero{}\szero{}} & 17 & $<$0.01\% \\
\texttt{I} & \texttt{1\szero{}1\szero{}} & 4 & 5.90\% & \texttt{.} & \texttt{\szero{}\szero{}\szero{}\szero{}\szero{}\szero{}\szero{}1\szero{}\szero{}1} & 11 & 0.05\% & \texttt{+} & \texttt{\szero{}\szero{}\szero{}\szero{}\szero{}\szero{}\szero{}\szero{}\szero{}\szero{}\szero{}1\szero{}111} & 16 & $<$0.01\% & \texttt{Ë} & \texttt{\szero{}\szero{}\szero{}\szero{}\szero{}\szero{}\szero{}\szero{}\szero{}\szero{}\szero{}\szero{}111\szero{}1} & 17 & $<$0.01\% \\
\texttt{S} & \texttt{1\szero{}11} & 4 & 5.41\% & \texttt{{\szero}} & \texttt{\szero{}\szero{}\szero{}\szero{}\szero{}\szero{}\szero{}1\szero{}1\szero{}} & 11 & 0.05\% & \texttt{\textbackslash{}n} & \texttt{\szero{}\szero{}\szero{}\szero{}\szero{}\szero{}\szero{}\szero{}\szero{}\szero{}\szero{}11\szero{}\szero{}\szero{}} & 16 & $<$0.01\% & \texttt{Ì} & \texttt{\szero{}\szero{}\szero{}\szero{}\szero{}\szero{}\szero{}\szero{}\szero{}\szero{}\szero{}\szero{}1111\szero{}} & 17 & $<$0.01\% \\
\texttt{R} & \texttt{11\szero{}\szero{}} & 4 & 5.00\% & \texttt{2} & \texttt{\szero{}\szero{}\szero{}\szero{}\szero{}\szero{}\szero{}1\szero{}11} & 11 & 0.05\% & \texttt{\%} & \texttt{\szero{}\szero{}\szero{}\szero{}\szero{}\szero{}\szero{}\szero{}\szero{}\szero{}\szero{}11\szero{}\szero{}1} & 16 & $<$0.01\% & \texttt{Í} & \texttt{\szero{}\szero{}\szero{}\szero{}\szero{}\szero{}\szero{}\szero{}\szero{}\szero{}\szero{}\szero{}11111} & 17 & $<$0.01\% \\
\texttt{H} & \texttt{11\szero{}1} & 4 & 4.76\% & \texttt{8} & \texttt{\szero{}\szero{}\szero{}\szero{}\szero{}\szero{}\szero{}11\szero{}\szero{}} & 11 & 0.04\% & \texttt{\&} & \texttt{\szero{}\szero{}\szero{}\szero{}\szero{}\szero{}\szero{}\szero{}\szero{}\szero{}\szero{}11\szero{}1\szero{}} & 16 & $<$0.01\% & \texttt{Î} & \texttt{\szero{}\szero{}\szero{}\szero{}\szero{}\szero{}\szero{}\szero{}\szero{}\szero{}\szero{}\szero{}\szero{}1\szero{}\szero{}\szero{}} & 17 & $<$0.01\% \\
\texttt{D} & \texttt{111\szero{}\szero{}} & 5 & 3.48\% & \texttt{3} & \texttt{\szero{}\szero{}\szero{}\szero{}\szero{}\szero{}\szero{}11\szero{}1} & 11 & 0.04\% & \texttt{$<$} & \texttt{\szero{}\szero{}\szero{}\szero{}\szero{}\szero{}\szero{}\szero{}\szero{}\szero{}\szero{}11\szero{}11} & 16 & $<$0.01\% & \texttt{Ï} & \texttt{\szero{}\szero{}\szero{}\szero{}\szero{}\szero{}\szero{}\szero{}\szero{}\szero{}\szero{}\szero{}\szero{}1\szero{}\szero{}1} & 17 & $<$0.01\% \\
\texttt{L} & \texttt{111\szero{}1} & 5 & 3.21\% & \texttt{4} & \texttt{\szero{}\szero{}\szero{}\szero{}\szero{}\szero{}\szero{}111\szero{}} & 11 & 0.04\% & \texttt{$>$} & \texttt{\szero{}\szero{}\szero{}\szero{}\szero{}\szero{}\szero{}\szero{}\szero{}\szero{}\szero{}111\szero{}\szero{}\szero{}} & 17 & $<$0.01\% & \texttt{Ð} & \texttt{\szero{}\szero{}\szero{}\szero{}\szero{}\szero{}\szero{}\szero{}\szero{}\szero{}\szero{}\szero{}\szero{}1\szero{}1\szero{}} & 17 & $<$0.01\% \\
\texttt{C} & \texttt{1111\szero{}} & 5 & 2.34\% & \texttt{5} & \texttt{\szero{}\szero{}\szero{}\szero{}\szero{}\szero{}\szero{}1111} & 11 & 0.04\% & \texttt{@} & \texttt{\szero{}\szero{}\szero{}\szero{}\szero{}\szero{}\szero{}\szero{}\szero{}\szero{}\szero{}111\szero{}\szero{}1} & 17 & $<$0.01\% & \texttt{Ñ} & \texttt{\szero{}\szero{}\szero{}\szero{}\szero{}\szero{}\szero{}\szero{}\szero{}\szero{}\szero{}\szero{}\szero{}1\szero{}11} & 17 & $<$0.01\% \\
\texttt{U} & \texttt{11111\szero{}} & 6 & 2.24\% & \texttt{9} & \texttt{\szero{}\szero{}\szero{}\szero{}\szero{}\szero{}\szero{}\szero{}1\szero{}\szero{}} & 11 & 0.03\% & \texttt{\textbackslash{}} & \texttt{\szero{}\szero{}\szero{}\szero{}\szero{}\szero{}\szero{}\szero{}\szero{}\szero{}\szero{}111\szero{}1\szero{}} & 17 & $<$0.01\% & \texttt{Ò} & \texttt{\szero{}\szero{}\szero{}\szero{}\szero{}\szero{}\szero{}\szero{}\szero{}\szero{}\szero{}\szero{}\szero{}11\szero{}\szero{}} & 17 & $<$0.01\% \\
\texttt{M} & \texttt{111111} & 6 & 2.05\% & \texttt{6} & \texttt{\szero{}\szero{}\szero{}\szero{}\szero{}\szero{}\szero{}\szero{}1\szero{}1} & 11 & 0.03\% & \texttt{\^{}} & \texttt{\szero{}\szero{}\szero{}\szero{}\szero{}\szero{}\szero{}\szero{}\szero{}\szero{}\szero{}111\szero{}11} & 17 & $<$0.01\% & \texttt{Ó} & \texttt{\szero{}\szero{}\szero{}\szero{}\szero{}\szero{}\szero{}\szero{}\szero{}\szero{}\szero{}\szero{}\szero{}11\szero{}1} & 17 & $<$0.01\% \\
\texttt{F} & \texttt{\szero{}\szero{}\szero{}1\szero{}\szero{}} & 6 & 1.95\% & \texttt{7} & \texttt{\szero{}\szero{}\szero{}\szero{}\szero{}\szero{}\szero{}\szero{}11\szero{}\szero{}} & 12 & 0.03\% & \texttt{`} & \texttt{\szero{}\szero{}\szero{}\szero{}\szero{}\szero{}\szero{}\szero{}\szero{}\szero{}\szero{}1111\szero{}\szero{}} & 17 & $<$0.01\% & \texttt{Ô} & \texttt{\szero{}\szero{}\szero{}\szero{}\szero{}\szero{}\szero{}\szero{}\szero{}\szero{}\szero{}\szero{}\szero{}111\szero{}} & 17 & $<$0.01\% \\
\texttt{W} & \texttt{\szero{}\szero{}\szero{}1\szero{}1} & 6 & 1.63\% & \texttt{\textquotedbl{}} & \texttt{\szero{}\szero{}\szero{}\szero{}\szero{}\szero{}\szero{}\szero{}11\szero{}1} & 12 & 0.03\% & \texttt{\{} & \texttt{\szero{}\szero{}\szero{}\szero{}\szero{}\szero{}\szero{}\szero{}\szero{}\szero{}\szero{}1111\szero{}1} & 17 & $<$0.01\% & \texttt{Õ} & \texttt{\szero{}\szero{}\szero{}\szero{}\szero{}\szero{}\szero{}\szero{}\szero{}\szero{}\szero{}\szero{}\szero{}1111} & 17 & $<$0.01\% \\
\texttt{P} & \texttt{\szero{}\szero{}\szero{}11\szero{}} & 6 & 1.60\% & \texttt{-} & \texttt{\szero{}\szero{}\szero{}\szero{}\szero{}\szero{}\szero{}\szero{}111\szero{}} & 12 & 0.03\% & \texttt{\}} & \texttt{\szero{}\szero{}\szero{}\szero{}\szero{}\szero{}\szero{}\szero{}\szero{}\szero{}\szero{}11111\szero{}} & 17 & $<$0.01\% & \texttt{Ö} & \texttt{\szero{}\szero{}\szero{}\szero{}\szero{}\szero{}\szero{}\szero{}\szero{}\szero{}\szero{}\szero{}\szero{}\szero{}1\szero{}\szero{}} & 17 & $<$0.01\% \\
\texttt{G} & \texttt{\szero{}\szero{}\szero{}111} & 6 & 1.56\% & \texttt{\textquotesingle{}} & \texttt{\szero{}\szero{}\szero{}\szero{}\szero{}\szero{}\szero{}\szero{}1111} & 12 & 0.02\% & \texttt{\~{}} & \texttt{\szero{}\szero{}\szero{}\szero{}\szero{}\szero{}\szero{}\szero{}\szero{}\szero{}\szero{}111111} & 17 & $<$0.01\% & \texttt{Ø} & \texttt{\szero{}\szero{}\szero{}\szero{}\szero{}\szero{}\szero{}\szero{}\szero{}\szero{}\szero{}\szero{}\szero{}\szero{}1\szero{}1} & 17 & $<$0.01\% \\
\texttt{Y} & \texttt{\szero{}\szero{}\szero{}\szero{}1\szero{}} & 6 & 1.46\% & \texttt{\_} & \texttt{\szero{}\szero{}\szero{}\szero{}\szero{}\szero{}\szero{}\szero{}\szero{}1\szero{}\szero{}} & 12 & 0.02\% & \texttt{¡} & \texttt{\szero{}\szero{}\szero{}\szero{}\szero{}\szero{}\szero{}\szero{}\szero{}\szero{}\szero{}\szero{}1\szero{}\szero{}\szero{}\szero{}} & 17 & $<$0.01\% & \texttt{Ù} & \texttt{\szero{}\szero{}\szero{}\szero{}\szero{}\szero{}\szero{}\szero{}\szero{}\szero{}\szero{}\szero{}\szero{}\szero{}11\szero{}} & 17 & $<$0.01\% \\
\texttt{B} & \texttt{\szero{}\szero{}\szero{}\szero{}11} & 6 & 1.18\% & \texttt{;} & \texttt{\szero{}\szero{}\szero{}\szero{}\szero{}\szero{}\szero{}\szero{}\szero{}1\szero{}1} & 12 & 0.02\% & \texttt{¿} & \texttt{\szero{}\szero{}\szero{}\szero{}\szero{}\szero{}\szero{}\szero{}\szero{}\szero{}\szero{}\szero{}1\szero{}\szero{}\szero{}1} & 17 & $<$0.01\% & \texttt{Ú} & \texttt{\szero{}\szero{}\szero{}\szero{}\szero{}\szero{}\szero{}\szero{}\szero{}\szero{}\szero{}\szero{}\szero{}\szero{}111} & 17 & $<$0.01\% \\
\texttt{V} & \texttt{\szero{}\szero{}\szero{}\szero{}\szero{}1\szero{}} & 7 & 0.85\% & \texttt{:} & \texttt{\szero{}\szero{}\szero{}\szero{}\szero{}\szero{}\szero{}\szero{}\szero{}11\szero{}} & 12 & 0.02\% & \texttt{À} & \texttt{\szero{}\szero{}\szero{}\szero{}\szero{}\szero{}\szero{}\szero{}\szero{}\szero{}\szero{}\szero{}1\szero{}\szero{}1\szero{}} & 17 & $<$0.01\% & \texttt{Û} & \texttt{\szero{}\szero{}\szero{}\szero{}\szero{}\szero{}\szero{}\szero{}\szero{}\szero{}\szero{}\szero{}\szero{}\szero{}\szero{}1\szero{}} & 17 & $<$0.01\% \\
\texttt{K} & \texttt{\szero{}\szero{}\szero{}\szero{}\szero{}11\szero{}} & 8 & 0.53\% & \texttt{=} & \texttt{\szero{}\szero{}\szero{}\szero{}\szero{}\szero{}\szero{}\szero{}\szero{}111\szero{}} & 13 & 0.02\% & \texttt{Á} & \texttt{\szero{}\szero{}\szero{}\szero{}\szero{}\szero{}\szero{}\szero{}\szero{}\szero{}\szero{}\szero{}1\szero{}\szero{}11} & 17 & $<$0.01\% & \texttt{Ü} & \texttt{\szero{}\szero{}\szero{}\szero{}\szero{}\szero{}\szero{}\szero{}\szero{}\szero{}\szero{}\szero{}\szero{}\szero{}\szero{}11} & 17 & $<$0.01\% \\
\texttt{{\ensuremath{\lightning}}} & \texttt{\szero{}\szero{}\szero{}\szero{}\szero{}111} & 8 & 0.34\% & \texttt{(} & \texttt{\szero{}\szero{}\szero{}\szero{}\szero{}\szero{}\szero{}\szero{}\szero{}1111} & 13 & 0.01\% & \texttt{Â} & \texttt{\szero{}\szero{}\szero{}\szero{}\szero{}\szero{}\szero{}\szero{}\szero{}\szero{}\szero{}\szero{}1\szero{}1\szero{}\szero{}} & 17 & $<$0.01\% & \texttt{Ý} & \texttt{\szero{}\szero{}\szero{}\szero{}\szero{}\szero{}\szero{}\szero{}\szero{}\szero{}\szero{}\szero{}\szero{}\szero{}\szero{}\szero{}1} & 17 & $<$0.01\% \\
\texttt{X} & \texttt{\szero{}\szero{}\szero{}\szero{}\szero{}\szero{}1\szero{}\szero{}} & 9 & 0.16\% & \texttt{)} & \texttt{\szero{}\szero{}\szero{}\szero{}\szero{}\szero{}\szero{}\szero{}\szero{}\szero{}1\szero{}\szero{}} & 13 & 0.01\% & \texttt{Ã} & \texttt{\szero{}\szero{}\szero{}\szero{}\szero{}\szero{}\szero{}\szero{}\szero{}\szero{}\szero{}\szero{}1\szero{}1\szero{}1} & 17 & $<$0.01\% & \texttt{Þ} & \texttt{\szero{}\szero{}\szero{}\szero{}\szero{}\szero{}\szero{}\szero{}\szero{}\szero{}\szero{}\szero{}\szero{}\szero{}\szero{}\szero{}\szero{}1} & 18 & $<$0.01\% \\
\texttt{J} & \texttt{\szero{}\szero{}\szero{}\szero{}\szero{}\szero{}1\szero{}1\szero{}} & 10 & 0.10\% & \texttt{\#} & \texttt{\szero{}\szero{}\szero{}\szero{}\szero{}\szero{}\szero{}\szero{}\szero{}\szero{}1\szero{}1} & 13 & 0.01\% & \texttt{Ä} & \texttt{\szero{}\szero{}\szero{}\szero{}\szero{}\szero{}\szero{}\szero{}\szero{}\szero{}\szero{}\szero{}1\szero{}11\szero{}} & 17 & $<$0.01\% & \texttt{[FILL]} & \texttt{\szero{}\szero{}\szero{}\szero{}\szero{}\szero{}\szero{}\szero{}\szero{}\szero{}\szero{}\szero{}\szero{}\szero{}\szero{}\szero{}\szero{}\szero{}} & 18 & $<$0.01\% \\
\bottomrule
\end{tabular}%
}

\end{table*}

\subsection{Protocol Boundary Conditions \& Canonical Precedence}
\label{sec:msg:limitations}

To eliminate ambiguity across station profiles, \LQ{} formalizes eight canonical precedence and boundary rules:
\begin{enumerate}[leftmargin=1.4em, itemsep=1.5pt plus 1pt minus 0.5pt, topsep=2pt, parsep=0pt]
  \item \textbf{Strict Canonical Precedence:} Transmitters \textbf{must} use the most compact format fitting caller and target: for CALL, Type~5 requires both stations standard without suffix; Type~6 applies if either has \texttt{/P} or target is non-standard; Type~7 if caller is non-standard. For replies, Type~8 precedes Type~11 and Type~9 precedes Types~10/12 if both are standard.
  \item \textbf{Deterministic Hash Disambiguation:} In Type~6, target hash matches against unsuffixed standard stations are rejected only if caller also lacks a suffix (Type~5 was mandatory); suffixed callers and Type~7 validly target standard stations. In 1-on-1 contacts, standard stations ignore hashed completions (Types~10, 11) from standard peers.
  \item \textbf{Universal Locator and SNR in Type~6:} Type~6 allocates 4 prefix bits and 24 target hash bits, leaving 49 bits for caller callsign, suffix, 15-bit locator, and 5-bit SNR ($4+24+28+1+15+5=77\text{b}$). Suffixed callers deliver locator and SNR on their initial call without extra steps.
  \item \textbf{Clear-Text Identity in Type~7:} Type~7 encodes the target via 20-bit hash ($H_{20} = H_{24} \gg 4$), the non-standard caller via clear-text 48-bit Base-38 representation ($\le 9$ characters), 1-bit suffix (\texttt{/P}), and 5-bit SNR ($3+20+48+1+5 = 77\text{b}$), delivering unhashed identity directly in Step~2.
  \item \textbf{Single-Target Multi-Station Framing:} When confirming one station in Fox mode (Types~11/12), Target~2 duplicates Target~1 (identical hash, and SNR in Type~11), decoded as a single station. In 1-on-1 contacts, Type~8 strictly supersedes Type~11; single-target Type~11 is reserved for non-standard callers or split F/H schedules.
  \item \textbf{Collision Resistance:} Uniform 24-bit callsign hashes ($2^{24} \approx 16.78\times 10^6$ bins) yield $P_{\text{collision}} \le 0.75\%$ across 500 stations, disambiguated via local callsign caches and canonical precedence.
  \item \textbf{Compound Callsigns:} Stations with callsigns exceeding 9 characters (up to 13 characters, e.g., \callsign{3B9/HB9IPH/P}, Category~D) utilize the 69-bit Base-38 field in CQ frames (Type~4); directed replies use CQ or free text (Type~13).
  \item \textbf{Verification \& Compatibility:} Bit-exact codec traces and golden vectors across all 13 message formats are verified in the reference suite~\cite{lqlib} across AWGN, Rayleigh fading, and clock-drift conditions. Compliant decoders treat unassigned opcodes (Types~14--16) as valid framing boundaries, safely ignoring them.
\end{enumerate}
\vspace*{-4.0pt}

\pagebreak

\section{Field Encodings}
\label{sec:fields}
{\setlength{\abovedisplayskip}{2.0pt plus 1.5pt minus 1.0pt}
\setlength{\belowdisplayskip}{2.0pt plus 1.5pt minus 1.0pt}

To maintain compatibility, \LQ{} adopts core field structures from FT8~\cite{ft8,wsjtx} (K9AN/K1JT)---28-bit callsigns, 15-bit grid locators, 20-bit CQ modifiers, and 5-bit signal reports---while introducing 1-bit portable suffixes, Base-38 non-standard callsigns, CRC-24/Q hashes, and a 104-character free-text Varicode.

% -------------------------------------------------------------------
\subsection{Standard Callsign (28 bits)}
\label{sec:std_call}
% -------------------------------------------------------------------

Adopted from FT8~\cite{ft8,wsjtx}, standard ITU amateur callsigns canonicalize to 6 characters anchored around decimal digit $d$ at position~3: $c_1 c_2 d s_1 s_2 s_3$. Prefix ($c_1, c_2$) right-aligns with space padding in $c_1$ if 1 character; suffix ($s_1, s_2, s_3$) left-aligns with spaces. Callsigns matching $c_1 c_2 d s_1 s_2 s_3$ without slashes are standard; all others are non-standard (Section~\ref{sec:nonstd_call}).

Index mappings: $c_1 \in \{\text{\textvisiblespace}, \text{0--9}, \text{A--Z}\}$ (radix 37: space $\to 0$, \texttt{0--9} $\to 1\dots 10$, \texttt{A--Z} $\to 11\dots 36$); $c_2 \in \{\text{0--9}, \text{A--Z}\}$ (radix 36: \texttt{0--9} $\to 0\dots 9$, \texttt{A--Z} $\to 10\dots 35$); $d \in \{\text{0--9}\}$ (radix 10); and $s_1, s_2, s_3 \in \{\text{\textvisiblespace}, \text{A--Z}\}$ (radix 27: space $\to 0$, \texttt{A--Z} $\to 1\dots 26$). With indices $(n_1, \dots, n_6)$, mixed-radix serialization packs into 28-bit integer $N_{28}$:
\begin{equation}
  N_{28} = (n_1 \cdot 36 \cdot 10 + n_2 \cdot 10 + n_3) \cdot 27^3 + n_4 \cdot 27^2 + n_5 \cdot 27 + n_6.
\label{eq:call28}
\end{equation}
Standard callsigns occupy $0 \le N_{28} \le N_{\max} = 262{,}177{,}559 < 2^{28}$; subsequent states encode tokens $\texttt{DE} = 262{,}177{,}560$, $\texttt{QRZ} = 262{,}177{,}561$, and $\texttt{CQ} = 262{,}177{,}562$. Decoders validate radix ranges to reject corrupted values.

% -------------------------------------------------------------------
\subsection{Standard Dedicated Suffixes (1 bit)}
\label{sec:std_portable_suffix}
% -------------------------------------------------------------------

To support field activations (POTA/SOTA) without hashing, frames reserve an explicit 1-bit suffix: \texttt{0}=\emph{none} (e.g., \callsign{HB9IPH}), \texttt{1}=\texttt{/P} (portable).

% -------------------------------------------------------------------
\subsection{Non-Standard Callsigns (Base-38 Big-Integer Packing)}
\label{sec:nonstd_call}
% -------------------------------------------------------------------

Non-standard or slashed callsigns pack as Base-38 strings over alphabet \textvisiblespace{} (0), \texttt{A}--\texttt{Z} (1--26), \texttt{/} (27), and \texttt{0}--\texttt{9} (28--37), right-padded with spaces to length $L$ and evaluated big-endian as $N_{\text{b38}} = \sum_{i=0}^{L-1} v_i \cdot 38^{L-1-i}$. \textbf{9-Char Callsigns (48 bits):} $38^9 < 2^{48}$ (Types 2, 3, 7, 10), supporting up to 9 characters plus 1-bit portable suffix (\texttt{/P}, e.g., \callsign{EA6/HB9IP/P}) or slashed calls (\callsign{HB9IPH/Q}). \textbf{13-Char Callsigns (69 bits):} $38^{13} < 2^{69}$ (Type 4), supporting compound callsigns up to 13 characters plus suffix in CQ (\callsign{3B9/HB9IPH/P}). Unnormalized whitespace is rejected.

\textbf{Alphabet Disambiguation:} This Base-38 alphabet (space $\to 0$, \texttt{A--Z} $\to 1\dots 26$, \texttt{/} $\to 27$, \texttt{0--9} $\to 28\dots 37$) places letters before digits, unlike the legacy WSJT-X order (\texttt{0--9} $\to 1\dots 10$, \texttt{A--Z} $\to 11\dots 36$) used in the 16-bit hash (Section~\ref{sec:hash}).

% -------------------------------------------------------------------
\subsection{Callsign Hashes (24-bit, 20-bit, and 16-bit) \& Dehashing}
\label{sec:hash}
% -------------------------------------------------------------------

Non-standard stations addressed in directed frames are identified via collision-resistant hashes:
\textbf{24-Bit Callsign Hash ($H_{24}$):} Computed over trimmed ASCII callsigns via CRC-24/Q~\cite{crc24q} ($G(x) = \texttt{0x1864CFB}$, init/final \texttt{0x000000}, non-reflected): $H_{24}(\text{call}) = \text{CRC-24/Q}(\text{call})$, yielding $2^{24} = 16{,}777{,}216$ states ($P_{\text{collision}} \le 0.75\%$ across 500 stations).
\textbf{20-Bit Target Hash ($H_{20}$):} Used in Type~7 directed replies as 20 MSBs: $H_{20}(\text{call}) = \lfloor H_{24}(\text{call}) / 2^4 \rfloor = H_{24}(\text{call}) \gg 4$, yielding $2^{20} = 1{,}048{,}576$ states (e.g., $H_{20} = \texttt{0x01A4F}$ for $H_{24} = \texttt{0x01A4F2}$), conserving 4 payload bits for reports while maintaining high collision resistance.

\textbf{16-Bit DX Callsign Hash ($H_{16}$):} Used in multi-station frames (Types 11 and 12), computed via the WSJT-X 16-bit multiplicative hash~\cite{wsjtx}. Callsigns right-pad to 11 characters over alphabet (\textvisiblespace{} $\to 0$, \texttt{0--9} $\to 1\dots 10$, \texttt{A--Z} $\to 11\dots 36$, \texttt{/} $\to 37$):
\begin{equation}
  H_{16}(\text{call}) = \lfloor (\textstyle\sum_{i=0}^{10} v_i \cdot 38^{10-i} \cdot 47055833459) / 2^{48} \rfloor \bmod 2^{16},
\end{equation}
yielding 65,536 states (e.g., $H_{16}(\text{\callsign{HB9IPH}}) = \texttt{0x7A8B}$); to prevent collisions, decoders restrict $H_{16}$ dehashing to actively called stations (evicted upon QSO completion or timeout).

\textbf{UI Resolution:} Resolved hashes display bracketed by callsign (e.g., \texttt{<HB9IPH>}); unresolved hashes display in hex (e.g., \texttt{<01A4F2>}, \texttt{<01A4F>}, or \texttt{<7A8B>}).

% -------------------------------------------------------------------
\subsection{Maidenhead Grid Locator (15 bits)}
\label{sec:locator}

Following the FT8 standard~\cite{ft8,wsjtx,maidenhead}, four-character Maidenhead grid locators ($F_1 F_2 D_1 D_2$, with fields $F_1, F_2 \in [\text{A--R}]$ and squares $D_1, D_2 \in [0\text{--}9]$) divide the globe into $18 \times 18 = 324$ fields of $10 \times 10 = 100$ squares ($32{,}400$ total squares). Packed into 15 bits ($N_{\text{grid}} < 32{,}400 < 2^{15} = 32{,}768$):
\begin{equation}
\begin{aligned}
  N_{\text{grid}} ={}& (F_1 - \text{'A'}) \cdot 1800 + (F_2 - \text{'A'}) \cdot 100 \\
                     & + (D_1 - \text{'0'}) \cdot 10 + (D_2 - \text{'0'}).
\end{aligned}
\label{eq:grid}
\end{equation}
Values $0 \le N_{\text{grid}} \le 32{,}399$ represent valid Maidenhead squares; $N_{\text{grid}} = 32{,}400$ is the blank locator sentinel; values $32{,}401 \le N_{\text{grid}} \le 32{,}767$ are rejected.

% -------------------------------------------------------------------
\subsection{CQ Modifiers (20 bits)}
\label{sec:modifier}
% -------------------------------------------------------------------

Directed CQ frames (Types~1 and 3) adopt the 20-bit modifier structure from WSJT-X / FT8~\cite{ft8,wsjtx} ($2^{20} = 1{,}048{,}576$ states):
\emph{1) Unmodified CQ:} $N_{\text{mod}} = 0$ designates general CQ calls.
\emph{2) 3-Digit Numeric (000--999):} $1 \le N_{\text{mod}} \le 1000$ conveys numeric designators (e.g., net/channel IDs \texttt{CQ 040}), mapped as $N_{\text{mod}} = \text{val} + 1$.
\emph{3) Alphanumeric Tokens ($\le 4$ chars):} $1001 \le N_{\text{mod}} \le 1{,}048{,}575$ encodes tokens up to 4 characters (e.g., \texttt{DX}, \texttt{NA}, \texttt{POTA}, \texttt{SOTA}, \texttt{QRP}, \texttt{CQWW}; values $> 2^{20}-1$ rejected). Tokens $< 4$ characters right-pad with spaces over a 32-symbol alphabet (\textvisiblespace{} $\to 0$, \texttt{A--Z} $\to 1\dots 26$, \texttt{0--4} $\to 27\dots 31$):
\begin{equation}
  N_{\text{mod}} = 1000 + \sum_{i=0}^{3} v_i \cdot 32^{3-i}.
\label{eq:modifier}
\end{equation}
Omitting digits 5--9 ensures $32^4 = 2^{20}$ fits without overflow; longer modifiers use 4-character abbreviations (\texttt{WPX}) or free text.

% -------------------------------------------------------------------
\subsection{Signal Reports (5 bits)}
\label{sec:rst}
% -------------------------------------------------------------------

Following FT8~\cite{ft8}, signal reports convey measured SNR from $-26$\,dB to $+5$\,dB via integer $N_{\text{snr}} = \text{clamp}(\text{SNR}_{\text{dB}} + 26, 0, 31) \in [0, 31]$. In directed \msgtype{CALL} replies (Types~5--7), reports display without prefix (e.g., \texttt{-03}); in acknowledgments (\msgtype{REPORT+73}, Types~8 and 11), \texttt{R} is prepended (e.g., \texttt{R-03}).

% -------------------------------------------------------------------
\subsection{Free-Text Varicode Encoding (73 bits)}
\label{sec:varicode}
% -------------------------------------------------------------------

Unstructured text in Type~13 (\msgtype{FREE TEXT}, 73 payload bits) uses a 104-character prefix code (Table~\ref{tab:varicode})~\cite{varicode} optimized from character frequencies~\cite{lewand}. The alphabet spans uppercase \texttt{A--Z}, digits \texttt{0--9}, space, 32 punctuation symbols, 32 Latin-1 characters (¡--Þ), newline \texttt{\textbackslash{}n}, \lightning{}, and \texttt{[FILL]}. Codewords range from 3 to 18 bits (mean 4.25\,bits/char), conveying \textbf{$\approx$17.2 characters} on average ($+32.3\%$ over conventional 13-character payloads). Unused bits are zero-padded; decoders terminate extraction at the first zero bit without emitting fill tokens.
}

\begin{figure*}[!t]
\centering
\vspace{-4pt}
\includegraphics[width=\textwidth]{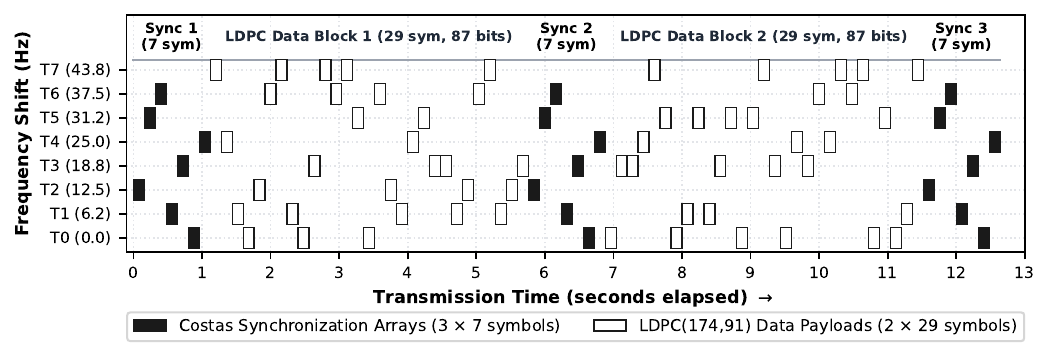}
\vspace{-6pt}
\caption{\LQ8 physical frame structure (8-GFSK, 50.0\,Hz bandwidth, 12.64\,s duration).}
\label{fig:waterfall_lq8}
\vspace{-2pt}
\end{figure*}

\pagebreak

\section{Physical Layer}
\label{sec:phy}

To ensure equal weak-signal sensitivity ($-21.0$\,dB SNR in 2500\,Hz bandwidth) and transceiver compatibility, \LQ8 deliberately adopts the core physical-layer modulation and error-correction coding of the FT8 protocol~\cite{ft8}, engineered by Steve Franke (K9AN) and Joe Taylor (K1JT).

\subsection{Channel Coding: CRC-14 and Systematic LDPC(174,91)}
\label{sec:crc14}
\label{sec:ldpc}

Link-layer encapsulation preserves FT8's channel-coding structure~\cite{ft8}:
(1)~\textbf{77-Bit Payload} from the application layer;
(2)~\textbf{14-Bit CRC Checksum} ($G_{\text{CRC14}}(x)$, Table~\ref{tab:crc14_params}) evaluated by shifting the 77-bit payload MSB-first into a 14-bit LFSR using FT8's generator polynomial \texttt{0x2757} followed by 14 trailing zeros, appending remainder ($d_{77}\dots d_{90}$) to form a 91-bit systematic block ($P_{\text{false}} \le 2^{-14} \approx 6.1\times 10^{-5}$); 5 zero bits pad to a 12-byte boundary; and
(3)~\textbf{Systematic LDPC(174,91) FEC} utilizing the identical Gallager generator matrix $G_{83\times 91}$ (Table~\ref{tab:ldpc_g})~\cite{ft8,gallager} producing 83 parity bits ($R \approx 0.523$, 174 codeword bits).

Systematic codeword $\mathbf{c} \in \mathbb{F}_2^{174}$ sets $c_j = d_j$ for $0 \le j < 91$. Parity bits are:
\begin{equation}
  c_{91+i} = \bigoplus_{j=0}^{90} G_{i, j} \cdot d_j, \quad 0 \le i < 83,
\end{equation}
where $G_{83 \times 91}$ is the Gallager matrix in Table~\ref{tab:ldpc_g} (23 hex digits/row, 91 MSB-aligned bits; bit 0 of digit 23 is zero pad). CRC-14 parameters are listed in Table~\ref{tab:crc14_params}.

\begin{table}[htbp]
\centering
\vspace{-3pt}
\caption{CRC-14 channel coding parameter specification.}
\label{tab:crc14_params}
\resizebox{\columnwidth}{!}{%
\renewcommand{\arraystretch}{0.92}%
\begin{tabular}{@{}ll@{}}
\toprule
\textbf{Parameter} & \textbf{Value / Specification} \\
\midrule
Generator Polynomial $G_{\text{CRC14}}(x)$ & $x^{14} + x^{13} + x^{10} + x^9 + x^8 + x^6 + x^4 + x^2 + x + 1$ \\
Hex / LFSR Representation & \texttt{0x2757} (14-bit: \texttt{10011101010111}$_2$; 15-bit monic: \texttt{0x6757}) \\
Initial State / Final XOR & \texttt{0x0000} (14 zeros) / \texttt{0x0000} (uninverted) \\
Input / Output Reflection & \texttt{False} / \texttt{False} (MSB-first evaluation \& output) \\
Bit Framing \& Padding & 77 payload bits + 14 CRC bits = 91 systematic bits ($+5$ pad $\to 96$) \\
Undetected Error Probability & $P_{\text{false}} \le 2^{-14} \approx 6.10 \times 10^{-5}$ \\
Invalid Checksum Sentinel & Remainder \texttt{0x0000} rejected as corrupt \\
Target FEC Input Block & Feeds LDPC$(174,91)$ generator $G_{83\times 91}$ \\
\bottomrule
\end{tabular}%
}
\vspace{-3pt}
\end{table}

\subsection{LQ8 Physical Transport \& UTC Slot Framing}
\label{sec:transport_profiles}
\label{sec:utc_sequence}

\LQ8 adopts FT8's continuous-phase 8-GFSK transport parameters~\cite{ft8}: 79 symbols in 12.64\,s within 15.0\,s UTC slots ($R_s = 6.25$\,Bd, $\Delta f = 6.25$\,Hz, 50.0\,Hz footprint, $-21.0$\,dB SNR sensitivity, Figure~\ref{fig:waterfall_lq8}). Transmissions occupy 3.0\,kHz USB segments with multiple concurrent subcarrier offsets.

For timestamp $t_{\text{UTC}}$, slot index and parity $s \in \{0, 1\}$ evaluate as $k = \lfloor t_{\text{UTC}} / T_{\text{slot}} \rfloor$ and $s = k \bmod 2$ (Even :00/:30 vs.\ Odd :15/:45). Guard interval $T_{\text{guard}} = T_{\text{slot}} - T_{\text{tx}}$ (2.36\,s in \LQ8, 2.46\,s in \LQ4) provides propagation and T/R turnaround margin; stations alternate parity across cycles.

\subsection{Modulation Mapping \& Continuous-Phase Synthesis}

The 3-bit Gray code maps channel bit triplets to tones: \texttt{000}$\to$0, \texttt{001}$\to$1, \texttt{011}$\to$2, \texttt{010}$\to$3, \texttt{110}$\to$4, \texttt{100}$\to$5, \texttt{101}$\to$6, and \texttt{111}$\to$7. Continuous-phase audio synthesis $s(t) = A(t) \cos(\phi(t))$ at $f_s = 12{,}000$\,Hz employs half-symbol raised-cosine envelope shaping ($T_{\text{ramp}} = T_s / 2 = 80.0$\,ms in \LQ8, $160.0$\,ms in \LQ16) and Gaussian frequency trajectories ($BT = 2.0$):
{\setlength{\abovedisplayskip}{2pt plus 1.5pt minus 1.0pt}\setlength{\belowdisplayskip}{2pt plus 1.5pt minus 1.0pt}
\begin{equation}
  \phi(t) = \phi_0 + 2\pi \int_0^t \left( f_0 + \sum_{k=0}^{N_{\text{sym}}-1} a_k \cdot \Delta f \cdot g(\tau - k T_s) \right) d\tau,
\end{equation}
where tone index $a_k \in \{0, 1, \dots, M-1\}$ is uncentered with tone 0 at base subcarrier $f_0$ ($\Delta f = 6.25$\,Hz in \LQ8). Frequency pulse $g(t)$ is truncated to duration $L = 3 T_s$ ($t \in [-1.5 T_s, +1.5 T_s]$):
\begin{equation}
  g(t) = \frac{1}{2 T_s} \left[ \operatorname{erf}\left( 2\pi B \frac{t + T_s/2}{\sqrt{2 \ln 2}} \right) - \operatorname{erf}\left( 2\pi B \frac{t - T_s/2}{\sqrt{2 \ln 2}} \right) \right].
\end{equation}
}

\subsection{Costas Synchronization \& DSP Pipeline}
\label{sec:dsp_pipeline}

While \LQ8 shares FT8's modulation, CRC-14, and LDPC codes, it deliberately departs in its synchronization structure to achieve physical cross-protocol isolation. Rather than FT8's sync sequence ($[3, 1, 4, 0, 6, 5, 2]$), 8-GFSK modes (\LQ8 and \LQ16) synchronize via three 7-symbol Costas arrays~\cite{costas} ($C_7 = [2, 5, 6, 1, 3, 0, 4]$) at symbols $0\text{--}6$, $36\text{--}42$, and $72\text{--}78$. The 174-bit systematic LDPC codeword is partitioned without bit interleaving into two 87-bit data blocks: Data Block~1 (symbols $7\text{--}35$, $29 \times 3 = 87$ bits) carries codeword bits $c_0\dots c_{86}$, while Data Block~2 (symbols $43\text{--}71$) carries $c_{87}\dots c_{173}$. Each symbol maps 3 consecutive bits MSB-first via the 3-bit Gray code. Preambles exhibit ideal autocorrelation with peak sidelobe $\le 1$ and $\le 1$ coincidence hit against FT8, preventing false decodes.

Receiver DSP executes the multi-pass decoding paradigm from WSJT-X~\cite{ft8,wsjtx}: sliding STFT (Hann window, 4-hop oversampling) generating spectral density $S[t_i, f_j]$, 2D matched filter detection across $\pm 0.5$\,s timing windows, 3-point parabolic frequency refinement ($|\Delta f_{\text{err}}| < 0.1$\,Hz), soft LLR extraction, log-domain LDPC belief-propagation decoding (up to 25 passes, terminating early upon $\mathbf{H}\cdot\hat{\mathbf{c}}^T = \mathbf{0}$ and CRC-14 verification), and time-frequency candidate subtraction to decode weaker signals.

% =============================================================================
% The LQ Digital Mode Family — Reference Implementation (lq_lib)
%
% Author:  Luis Quesada (HB9IPH)
% Web:     https://luisquesada.com
% Portal:  https://lquesada.github.io/lq_lib/
% GitHub:  https://github.com/lquesada/lq_lib
% App:     qFT8 — Portable Amateur Radio (https://qft8.com)
%
% License: MIT License (https://github.com/lquesada/lq_lib/blob/main/LICENSE)
%
% Copyright (c) 2026 Luis Quesada (HB9IPH)
%
% Permission is hereby granted, free of charge, to any person obtaining a copy
% of this software and associated documentation files (the "Software"), to deal
% in the Software without restriction, including without limitation the rights
% to use, copy, modify, merge, publish, distribute, sublicense, and/or sell
% copies of the Software, and to permit persons to whom the Software is
% furnished to do so, subject to the following conditions:
%
% The above copyright notice and this permission notice shall be included in
% all copies or substantial portions of the Software.
%
% THE SOFTWARE IS PROVIDED "AS IS", WITHOUT WARRANTY OF ANY KIND, EXPRESS OR
% IMPLIED, INCLUDING BUT NOT LIMITED TO THE WARRANTIES OF MERCHANTABILITY,
% FITNESS FOR A PARTICULAR PURPOSE AND NONINFRINGEMENT. IN NO EVENT SHALL THE
% AUTHORS OR COPYRIGHT HOLDERS BE LIABLE FOR ANY CLAIM, DAMAGES OR OTHER
% LIABILITY, WHETHER IN AN ACTION OF CONTRACT, TORT OR OTHERWISE, ARISING FROM,
% OUT OF OR IN CONNECTION WITH THE SOFTWARE OR THE USE OR OTHER DEALINGS IN THE
% SOFTWARE.
% =============================================================================

\begin{table*}[t]
\centering
\small
\vspace{-4pt}
\setlength{\tabcolsep}{6pt}
\renewcommand{\arraystretch}{0.95}
\caption{The complete Gallager $\text{LDPC}(174, 91)$ generator matrix $G_{83 \times 91}$ (83 rows $\times$ 23 hex digits). Each row represents 92 bits MSB-aligned; bit 0 of the 23rd hex digit is an unused zero padding bit.}
\label{tab:ldpc_g}
\vspace{-2pt}
\begin{tabular}{@{}cl | cl | cl@{}}
\toprule
\textbf{Row} & \textbf{Generator Vector (23 Hex Digits)} & \textbf{Row} & \textbf{Generator Vector (23 Hex Digits)} & \textbf{Row} & \textbf{Generator Vector (23 Hex Digits)} \\
\midrule
00 & \texttt{8329ce11bf31eaf509f27fc} & 28 & \texttt{8eba1a13db3390bd6718cec} & 56 & \texttt{7ee18230c583cccc57d4b08} \\
01 & \texttt{761c264e25c259335493132} & 29 & \texttt{753844673a27782cc42012e} & 57 & \texttt{a066cb2fedafc9f52664126} \\
02 & \texttt{dc265902fb277c6410a1bdc} & 30 & \texttt{06ff83a145c37035a5c1268} & 58 & \texttt{bb23725abc47cc5f4cc4cd2} \\
03 & \texttt{1b3f417858cd2dd33ec7f62} & 31 & \texttt{3b37417858cc2dd33ec3f62} & 59 & \texttt{ded9dba3bee40c59b5609b4} \\
04 & \texttt{09fda4fee04195fd034783a} & 32 & \texttt{9a4a5a28ee17ca9c324842c} & 60 & \texttt{d9a7016ac653e6decdc9036} \\
05 & \texttt{077cccc11b8873ed5c3d48a} & 33 & \texttt{bc29f465309c977e89610a4} & 61 & \texttt{9ad46aed5f707f280ab5fc4} \\
06 & \texttt{29b62afe3ca036f4fe1a9da} & 34 & \texttt{2663ae6ddf8b5ce2bb29488} & 62 & \texttt{e5921c77822587316d7d3c2} \\
07 & \texttt{6054faf5f35d96d3b0c8c3e} & 35 & \texttt{46f231efe457034c1814418} & 63 & \texttt{4f14da8242a8b86dca73352} \\
08 & \texttt{e20798e4310eed27884ae90} & 36 & \texttt{3fb2ce85abe9b0c72e06fbe} & 64 & \texttt{8b8b507ad467d4441df770e} \\
09 & \texttt{775c9c08e80e26ddae56318} & 37 & \texttt{de87481f282c153971a0a2e} & 65 & \texttt{22831c9cf1169467ad04b68} \\
10 & \texttt{b0b811028c2bf997213487c} & 38 & \texttt{fcd7ccf23c69fa99bba1412} & 66 & \texttt{213b838fe2ae54c38ee7180} \\
11 & \texttt{18a0c9231fc60adf5c5ea32} & 39 & \texttt{f0261447e9490ca8e474cec} & 67 & \texttt{5d926b6dd71f085181a4e12} \\
12 & \texttt{76471e8302a0721e01b12b8} & 40 & \texttt{4410115818196f95cdd7012} & 68 & \texttt{66ab79d4b29ee6e69509e56} \\
13 & \texttt{ffbccb80ca8341fafb47b2e} & 41 & \texttt{088fc31df4bfbde2a4eafb4} & 69 & \texttt{958148682d748a38dd68baa} \\
14 & \texttt{66a72a158f9325a2bf67170} & 42 & \texttt{b8fef1b6307729fb0a078c0} & 70 & \texttt{b8ce020cf069c32a723ab14} \\
15 & \texttt{c4243689fe85b1c51363a18} & 43 & \texttt{5afea7acccb77bbc9d99a90} & 71 & \texttt{f4331d6d461607e95752746} \\
16 & \texttt{0dff739414d1a1b34b1c270} & 44 & \texttt{49a7016ac653f65ecdc9076} & 72 & \texttt{6da23ba424b9596133cf9c8} \\
17 & \texttt{15b48830636c8b99894972e} & 45 & \texttt{1944d085be4e7da8d6cc7d0} & 73 & \texttt{a636bcbc7b30c5fbeae67fe} \\
18 & \texttt{29a89c0d3de81d665489b0e} & 46 & \texttt{251f62adc4032f0ee714002} & 74 & \texttt{5cb0d86a07df654a9089a20} \\
19 & \texttt{4f126f37fa51cbe61bd6b94} & 47 & \texttt{56471f8702a0721e00b12b8} & 75 & \texttt{f11f106848780fc9ecdd80a} \\
20 & \texttt{99c47239d0d97d3c84e0940} & 48 & \texttt{2b8e4923f2dd51e2d537fa0} & 76 & \texttt{1fbb5364fb8d2c9d730d5ba} \\
21 & \texttt{1919b75119765621bb4f1e8} & 49 & \texttt{6b550a40a66f4755de95c26} & 77 & \texttt{fcb86bc70a50c9d02a5d034} \\
22 & \texttt{09db12d731faee0b86df6b8} & 50 & \texttt{a18ad28d4e27fe92a4f6c84} & 78 & \texttt{a534433029eac15f322e34c} \\
23 & \texttt{488fc33df43fbdeea4eafb4} & 51 & \texttt{10c2e586388cb82a3d80758} & 79 & \texttt{c989d9c7c3d3b8c55d75130} \\
24 & \texttt{827423ee40b675f756eb5fe} & 52 & \texttt{ef34a41817ee02133db2eb0} & 80 & \texttt{7bb38b2f0186d46643ae962} \\
25 & \texttt{abe197c484cb74757144a9a} & 53 & \texttt{7e9c0c54325a9c15836e000} & 81 & \texttt{2644ebadeb44b9467d1f42c} \\
26 & \texttt{2b500e4bc0ec5a6d2bdbdd0} & 54 & \texttt{3693e572d1fde4cdf079e86} & 82 & \texttt{608cc857594bfbb55d69600} \\
27 & \texttt{c474aa53d70218761669360} & 55 & \texttt{bfb2cec5abe1b0c72e07fbe} &    & \\
\bottomrule
\end{tabular}
\end{table*}

\pagebreak

\section{Extended Modes: LQ16, LQ4, and LQ2}
\label{sec:extended_modes}

While \textbf{LQ8} is the primary mode for 15-second scheduled weak-signal communications, the 77-bit variable-length architecture scales across three complementary physical profiles (Table~\ref{tab:comparison}):
\begin{itemize}[leftmargin=1.2em, topsep=1.5pt plus 1.5pt minus 0.5pt, itemsep=0.8pt plus 1.2pt minus 0.4pt, parsep=0pt]
  \item \textbf{LQ16 (Slow / Ultra Weak):} 8-GFSK mode for severe weak-signal conditions ($-24.0$\,dB SNR threshold in 2500\,Hz) with 320\,ms symbol period (25.0\,Hz bandwidth, $BT = 2.0$, 25.28\,s duration) for extreme weak-signal and EME paths.
  \item \textbf{LQ4 (Fast / Contest):} 4-GFSK mode for rapid contesting and pileups (30.0\,s contact, $120$\,QSOs/h standard, $480$\,QSOs/h Fox/Hound) with 48\,ms symbol period, 83.3\,Hz bandwidth, and 5.04\,s transmission in 7.5\,s UTC slots.
  \item \textbf{LQ2 (Turbo Burst):} 4-GFSK high-throughput profile for dynamic burst channels (15.0\,s full contact, $240$\,QSOs/h standard, $960$\,QSOs/h Fox/Hound) with 24\,ms symbol period, 166.7\,Hz bandwidth, and 2.52\,s transmission in 3.75\,s slots for meteor scatter.
\end{itemize}

\subsection{4-GFSK Frame Structure \& Modulation}

\LQ4 and \LQ2 transmit 105 symbols/frame using continuous-phase 4-GFSK ($BT = 1.0$, tone spacing $\Delta f = 1/T_s$, modulation index $h = 1.0$):
\begin{itemize}[leftmargin=1.2em, topsep=1.5pt plus 1.5pt minus 0.5pt, itemsep=0.8pt plus 1.2pt minus 0.4pt, parsep=0pt]
  \item \textbf{Consolidated Framing Sequence:} The 105-symbol frame is partitioned into: tone-0 ramp-up (sym~0, $T_{\text{ramp}} = T_s / 2$, 24\,ms in \LQ4, 12\,ms in \LQ2); Costas sync $S_1 = [1, 3, 2, 0]$ (syms~1--4); Data Block~1 (syms~5--33, 29 symbols, codeword bits $0\dots 57$); Costas $S_2 = [3, 1, 0, 2]$ (syms~34--37); Data Block~2 (syms~38--66, 29 symbols, bits $58\dots 115$); Costas $S_3 = [0, 2, 3, 1]$ (syms~67--70); Data Block~3 (syms~71--99, 29 symbols, bits $116\dots 173$); Costas $S_4 = [2, 0, 1, 3]$ (syms~100--103); and tone-0 ramp-down (sym~104, $T_{\text{ramp}} = T_s / 2$).
  \item \textbf{Scrambling \& Whitening:} Prior to CRC-14/LDPC channel encoding, the 77-bit payload is bitwise XORed with the 10-byte whitening sequence (\texttt{4A 5E 89 B4 B0 8A 79 55 BE 28}$_{16}$) to suppress discrete spectral lines. Bits 77--79 (byte 9, \texttt{0x28}) are zeros prior to CRC-14 calculation.
  \item \textbf{Data Framing \& Gray Coding:} Codeword bits are mapped sequentially without bit interleaving. The three 29-symbol data blocks map consecutive pairs $(b_{2k}, b_{2k+1})$ of the 174-bit codeword to 4-GFSK tones via Gray coding (\texttt{00}$\to$0, \texttt{01}$\to$1, \texttt{11}$\to$2, \texttt{10}$\to$3).
  \item \textbf{Ramping \& Guard Margins:} Tone-0 symbols at indices 0 and 104 frame the sequence ($5.04$\,s for \LQ4, $2.52$\,s for \LQ2). Guard intervals ($2.46$\,s in \LQ4, $1.23$\,s in \LQ2) absorb propagation delay and T/R turnaround.
\end{itemize}

\subsection{Profile Selection and Channel Matching}
\label{sec:channel_matching}

Transceivers dynamically select operational profiles according to ionospheric channel coherence time $T_c$ and link signal-to-noise ratio:
\begin{itemize}[leftmargin=1.2em, topsep=1.5pt plus 1.5pt minus 0.5pt, itemsep=0.8pt plus 1.2pt minus 0.4pt, parsep=0pt]
  \item When $\text{SNR} < -21.0$\,dB and Doppler spread $\sigma_f \le 0.5$\,Hz, \LQ16 provides a $+3.0$\,dB gain over \LQ8, sustaining readability down to $-24.0$\,dB for extreme paths.
  \item In rapid-fading channels ($T_c < 10$\,s), \LQ4's 48\,ms symbols and 7.5\,s slots ensure robust sync, completing verified contacts in 30.0\,s before phase degrades.
  \item For dynamic burst channels (e.g., meteor scatter), \LQ2's 2.52\,s bursts capture transient trails, completing 4-slot exchanges in 15.0\,s.
\end{itemize}
\vspace*{-4.0pt}

\begin{table*}[!t]
\centering
\caption{Multi-mode parameter and performance comparison across standard reference modes and the \LQ{} family.}
\label{tab:comparison}
\small
\setlength{\tabcolsep}{5pt}
\renewcommand{\arraystretch}{0.86}
\begin{tabular*}{\textwidth}{@{\extracolsep{\fill}}lcccccc@{}}
\toprule
\textbf{Parameter} & \textbf{FT8 (Ref)} & \textbf{FT4 (Ref)} & \textbf{\LQ8} & \textbf{\LQ16} & \textbf{\LQ4} & \textbf{\LQ2} \\
\midrule
Modulation            & 8-GFSK   & 4-GFSK   & 8-GFSK   & 8-GFSK   & 4-GFSK   & 4-GFSK \\
Bandwidth             & 50.0\,Hz & 83.3\,Hz & 50.0\,Hz & 25.0\,Hz & 83.3\,Hz & 166.7\,Hz \\
Time slot             & 15.0\,s  & 7.5\,s   & 15.0\,s  & 30.0\,s  & 7.5\,s   & 3.75\,s \\
Transmission time     & 12.64\,s & 5.04\,s  & 12.64\,s & 25.28\,s & 5.04\,s  & 2.52\,s \\
Sensitivity (SNR)     & $-$21.0\,dB& $-$17.5\,dB & $-$21.0\,dB& $-$24.0\,dB & $-$17.5\,dB & $-$14.0\,dB \\
Relative RF power req.& \textbf{0\,dB (ref)}& +3.5\,dB (2.2$\times$)& 0\,dB (ident.)& $-$3.0\,dB (0.5$\times$)& +3.5\,dB (2.2$\times$)& +7.0\,dB (5.0$\times$) \\
\midrule
Type coding           & Fixed    & Fixed    & Huffman  & Huffman  & Huffman  & Huffman \\
QSO messages          & 5--6     & 5--6     & 4        & 4        & 4        & 4 \\
QSO duration          & 75.0--90.0\,s & 37.5--45.0\,s & 60.0\,s & 120.0\,s & 30.0\,s & 15.0\,s \\
Standard QSO Rate$^*$ & 40--48/h & 80--96/h & 60/h     & 30/h     & 120/h    & 240/h \\
Pileup Rate (1-carrier)$^*$& 60/h & 120/h   & 240/h    & 120/h    & 480/h    & 960/h \\
Pileup Rate (2-carrier)$^*$& 120/h& 240/h   & 480/h    & 240/h    & 960/h    & 1920/h \\
\midrule
Standard Suffixes     & /P, /R   & /P, /R   & /P       & /P       & /P       & /P       \\
Max non-std callsign  & 11 chars & 11 chars & 13 + std suffix & 13 + std suffix & 13 + std suffix & 13 + std suffix \\
Callsign hash length  & 10/12/22-b& 10/12/22-b& 24/20/16-b& 24/20/16-b& 24/20/16-b& 24/20/16-b \\
Free-text alphabet    & 42 chars & 42 chars & 104 chars & 104 chars & 104 chars & 104 chars \\
Free-text capacity    & 71 bits ($\sim$13 c) & 71 bits ($\sim$13 c) & 73 bits ($\sim$17 c) & 73 bits ($\sim$17 c) & 73 bits ($\sim$17 c) & 73 bits ($\sim$17 c) \\
\bottomrule
\multicolumn{7}{@{}p{\textwidth}@{}}{\footnotesize $^*$FT8/FT4 contacts require 5--6 messages (75--90\,s / 37.5--45\,s) using \texttt{RR73} or \texttt{RRR}/73; \LQ{} completes contacts in 4 messages (60\,s in \LQ8). Peak rates assume continuous zero-loss operation. Pipelined pileup rates reflect F/H mode confirming 1 station/cycle in FT8/FT4 vs.\ 2 in \LQ{} (Type~11).}\end{tabular*}
\end{table*}

\pagebreak

\section{Conclusion}
\label{sec:conclusion}

\textbf{LQ8} demonstrates that continuous-phase 8-GFSK and systematic LDPC$(174,91)$ transport deliver higher throughput through variable-length prefix framing while maintaining full backward physical compatibility and $-21.0$\,dB SNR sensitivity in a 2500\,Hz reference bandwidth. By optimizing message prefix allocations to match operational entropy, \LQ8 packs callsigns, locators, and reports into a 77-bit budget, finalizing confirmed two-way contacts in 4 transmission slots (60\,s). For high-density pileups, \LQ8 introduces the single-carrier \msgtype{MULTI-REPORT+73} protocol, confirming two answering stations simultaneously without RF power-splitting penalties and achieving throughputs up to 240\,QSOs/h (480\,QSOs/h in dual-carrier operation).

The unified protocol family extends this architecture across three complementary physical profiles. \LQ16 scales symbol duration to 320\,ms, delivering a $+3.0$\,dB sensitivity advantage ($-24.0$\,dB SNR threshold) across ultra-weak fading and EME paths. For rapid contesting and dynamic propagation, \LQ4 (7.5\,s slots) and \LQ2 (3.75\,s slots) utilize 4-GFSK modulation and Costas sync arrays to complete contacts in 30.0\,s and 15.0\,s, supporting burst rates up to 960\,QSOs/h on VHF/UHF meteor scatter and sporadic-E openings.

The complete software suite is implemented in the open-source C++ reference library (\texttt{lq\_lib}) under the MIT License, featuring an automated verification harness achieving $>90\%$ line coverage alongside comprehensive golden vector test suites across all operational profiles. End-to-end field verification has been demonstrated in real-world portable activations via \textbf{qFT8} across HF bands. Bit-exact codec traces for all 13 message formats are verified under AWGN, Rayleigh fading, and clock drift; reference sensitivities span $-14.0$\,dB (\LQ2) to $-24.0$\,dB (\LQ16).

\section*{Acknowledgment}
The original concept, protocol architecture, and reference implementations were developed by the author. Generative AI was used to assist with background research, code testing, and manuscript drafting. The author thanks the global amateur radio community for valuable feedback during early on-air testing.

\pagebreak

\end{document}